\documentclass[twocolumn,prc,showpacs,preprintnumbers,superscriptaddress,floatfix]{revtex4}
\usepackage{dcolumn}
\usepackage{bm}
\usepackage{longtable}
\usepackage{mathrsfs}
\usepackage{graphicx,epsfig,latexsym,amssymb}
\usepackage{multirow,amsmath,array,booktabs,color}
\usepackage[section]{placeins}

\begin{document}

\title{Probing the resonance of Dirac particle by the application of complex momentum representation}
\author{Niu Li}
\affiliation{School of Physics and Materials Science, Anhui University, Hefei 230601,
P.R.China}
\author{Min Shi}
\affiliation{School of Physics and Materials Science, Anhui University, Hefei 230601,
P.R.China}
\author{Jian-You Guo}
\email[E-mail:]{jianyou@ahu.edu.cn}
\affiliation{School of Physics and Materials Science, Anhui University, Hefei 230601,
P.R.China}
\author{Zhong-Ming Niu}
\email[E-mail:]{zmniu@ahu.edu.cn}
\affiliation{School of Physics and Materials Science, Anhui University, Hefei 230601,
P.R.China}
\author{Haozhao Liang}
\email[E-mail:]{haozhao.liang@riken.jp}
\affiliation{RIKEN Nishina Center, Wako 351-0198, Japan}
\affiliation{Department of Physics, Graduate School of Science, The University of Tokyo, Tokyo 113-0033, Japan}
\date{\today }

\begin{abstract}
Resonance plays critical roles in the formation of many physical
phenomena, and several methods have been developed for the exploration of
resonance. In this work, we propose a new scheme for resonance by solving
the Dirac equation in complex momentum representation, in which the resonant
states are exposed clearly in complex momentum plane and the resonance
parameters can be determined precisely without imposing unphysical
parameters. Combining with the relativistic mean-field theory, this method
is applied to probe the resonances in $^{120}$Sn with the energies,
widths, and wavefunctions being obtained. Comparing with other methods, this method is not only very effective for narrow resonances, but also can be reliably applied to broad resonances.
\end{abstract}

\pacs{21.60.Jz,21.10.Pc,25.70.Ef}
\maketitle


Resonance is the most striking phenomenon in the whole range of
scattering experiments, and appears widely in atomic, molecular, and nuclear
physics~\cite{Taylor72} as well as in chemical reactions~\cite{Yang15}.
Resonance plays a critical role in the formation of many physical phenomena,
such as halo~\cite{Meng96}, giant halo~\cite{Meng98,YZhang12}, deformed halo~%
\cite{Zhou10,Hamamoto10}, and quantum halos~\cite{Jensen04}.
The contribution of the continuum to
the giant resonance mainly comes from single-particle resonance~\cite%
{Curutchet89,Cao02}. Moreover, resonance is also closely relevant to the
nucleosynthesis of chemical elements in the universe~\cite%
{SZhang12,Faestermann15}. Therefore, study on the resonance is one of the
hottest topics in different fields of physics.

Based on the conventional scattering theory, a
series of methods for resonance have been proposed, such as \emph{R}-matrix method~\cite%
{Wigner47,Hale87}, \emph{K}-matrix method~\cite{Humblet91}, \emph{S}-matrix
method~\cite{Taylor72,Cao02}, Jost function approach~\cite{Lu12,Lu13}, Green's
function method~\cite{Economou06,Sun14}, and so on. These methods have gained
success in handling resonant and scattering problems. Unfortunately,
solution of the scattering problem turns out to be a very difficult task
both from the formal as well as from the computational points of view. For
this reason, several bound-state-like methods have been developed, which include
the real stabilization method (RSM)~\cite{Hazi70}, the analytic continuation
in the coupling constant (ACCC) approach~\cite{Kukulin89}, and the complex
scaling method (CSM)~\cite{Moiseyev98}.

The RSM, which identifies the resonant states in terms of independence of
the calculated results on model parameters, is simple and able to provide rough
results for the resonance parameters. For better results, many efforts have
been made in improving the RSM calculations~\cite{Taylor76,Mandelshtam94}.
The application of RSM to the relativistic mean-field (RMF) theory has been
developed in Ref.~\cite{LZhang08}. In the ACCC approach, a resonant state
is processed as an analytic continuation of bound state, which is direct and
convenient in determining the resonance parameters~\cite{Kukulin89}.
Presently, this approach has been combined with the RMF theory~\cite%
{Yang01,Zhang04}, and some exotic phenomena in nuclei were well explained~%
\cite{Guo06,Xu15}. The CSM is one of the most frequently used methods for
exploration of the resonance in atomic and molecular physics~\cite%
{Moiseyev98} as well as in nuclear physics~\cite%
{Michel09,Kruppa14,Carbonell14,Myo14,Papadimitriou15}. Recently, we have
applied the CSM to the relativistic framework~\cite{Guo101}, and obtained
satisfactory descriptions of the single-particle resonances in spherical
nuclei~\cite{Guo102,Guo103,Shi15} and deformed nuclei~\cite{Shi14}.

Although these bound-state-like methods are efficient in handling the unbound
problems, there still exist some shortcomings. The RSM is simple but not
precise enough for determining the resonance parameters, and it is often
used as a first step for other methods. There exists certain dependence
of the calculated results on the range of analytical continuation and the
order of polynomial in Pad\'e approximation in the ACCC calculations. The
CSM calculations are quite accurate, but it is not completely independent on the
rotation angle in the actual calculations with a finite basis. Hence, it is
necessary to explore new schemes without introducing any unphysical
parameters, but able to obtain accurately the concerned results.

From the scattering theory, it is known that the bound states populate on
the imaginary axis in the momentum plane, while the resonances locate at the
fourth quadrant. If we devise a scheme to solve directly the equation of
motion in the complex momentum space, we can obtain not only the bound states but
also resonant states. There have been some investigations on how to obtain
the bound states~\cite{Sukumar79,Kwon78} and resonant states~\cite{Berggren68, Hagen06,Deltuva15} using momentum representation in nonrelativistic case, and used later as ``Berggren representation" in shell model
calculations~\cite{Liotta96,Michel02}. However, probing the resonance of Dirac particle by the application of complex momentum representation is still missing. As the resonance of Dirac particle is widely concerned in many fields~\cite{Shah16,Zhou14,Stefanska16,Schulz15} and almost all the methods have been developed for description of the relativistic resonance~\cite{LZhang08,Zhang04,Guo101,Grineviciute12,Fuda01,Horodecki00,Sun14}, in this paper we will establish a new scheme for the resonance of Dirac particle using the complex momentum representation, which can be implemented straightforwardly to the relevant studies in other fields. In the following, we first introduce the theoretical formalism.

Considering that the RMF theory is very successful in describing nuclear phenomena~\cite{Serot86,Ring96,Vretenar05,Meng06,Niksic11,Liang15,Meng15}
and astrophysics phenomena~\cite{Sun08,Niu09,Meng11,Xu13,Niu13,Niu13b}, without loss of generality, we explore the resonance of Dirac particle based on the RMF framework, where the Dirac equation describing nucleons can be written as
\begin{equation}
\left[ \vec{\alpha}\cdot \vec{p}+\beta \left( M+S\right) +V\right] \psi
=\varepsilon \psi ,  \label{Diraceq1}
\end{equation}%
where $M$ represents the nucleon mass, $\vec{\alpha }$ and $\beta $ are the Dirac matrices, $S$ and $V$ are
the scalar and vector potentials, respectively. The details of the RMF
theory can refer to the literatures~\cite{Serot86,Ring96,Vretenar05,Meng06}%
. The solutions of Eq.~(\ref{Diraceq1}) include the bound states, resonant
states, and non-resonant continuum. The bound states can be obtained with
conventional methods. For the resonant states, several techniques have
been developed, but there exist some shortcomings in these methods. Here, we
will establish a new scheme for the resonances by solving the Dirac equation
using complex momentum representation.

The wavefunction of a free particle with momentum $\vec p$ or wavevector $\vec{k}%
= $ $\vec{p}/\hbar $ is denoted as $\left\vert \vec{k}\right\rangle $. In
the momentum representation, the Dirac equation (\ref{Diraceq1}) can be
expressed as
\begin{equation}
\int d\vec{k}^{\prime }\left\langle \vec{k}\right\vert H\left\vert \vec{k}%
^{\prime }\right\rangle \psi \left( \vec{k}^{\prime }\right) =\varepsilon
\psi \left( \vec{k}\right) ,  \label{Diraceq2}
\end{equation}%
where $H=\vec{\alpha}\cdot \vec{p}+\beta \left( M+S\right) +V$ is the Dirac
Hamiltonian, and $\psi \left( \vec{k}\right) $ is the momentum wavefunction.
For a spherical system, $\psi \left( \vec{k}\right) $ can be split into the
radial and angular parts as:$\ \ \ \ \ \ \ $%
\begin{equation}
\psi \left( \vec{k}\right) =\left(
\begin{array}{c}
f(k)\phi _{ljm_{j}}\left( \Omega _{k}\right) \\
g(k)\phi _{\tilde{l}jm_{j}}\left( \Omega _{k}\right)%
\end{array}%
\right) ,  \label{wavefunction}
\end{equation}%
where $\phi _{ljm_{j}}\left( \Omega _{k}\right) $ is a two-dimensional
spinor $\phi _{ljm_{j}}\left( \Omega _{k}\right) =\left[ \chi _{1/2}\left(
s\right) \otimes Y_{l}\left( \Omega _{k}\right) \right] _{jm_{j}}$. Quantum number $l$ ($%
\tilde{l}$) is the orbital angular momentum corresponding to
the large (small) component of Dirac spinor. The relationship between $l$ and $%
\tilde{l}$ is related to the total angular momentum quantum number $j$ with $%
\tilde{l}=2j-l$.

\begin{figure}
\includegraphics[width=8.5cm]{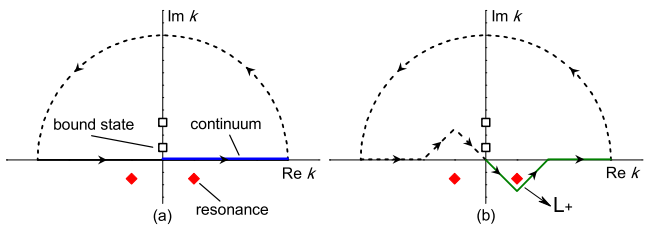}
\caption{(Color online) The complex momentum plane. The black open squares
represent the bound states and the red filled diamonds resonant states with the
mirroring of the states in the imaginary axis.}
\label{Fig1}
\end{figure}

Putting the wavefunction (\ref{wavefunction}) into Eq.~(\ref%
{Diraceq2}), the Dirac equation is reduced to the following form:
\begin{equation}
\left\{
\begin{array}{c}
Mf(k)-kg(k)+\int k^{\prime 2}dk^{\prime }V^{+}\left( k,k^{\prime }\right)
f(k^{\prime })=\varepsilon f(k), \\
-kf(k)-Mg(k)+\int k^{\prime 2}dk^{\prime }V^{-}\left( k,k^{\prime }\right)
g(k^{\prime })=\varepsilon g(k),%
\end{array}%
\right.   \label{Diraceq3}
\end{equation}%
with
\begin{align}
V^{+}\left( k,k^{\prime }\right)  &=\frac{2}{\pi }\int r^{2}dr\left[
V\left( r\right) +S\left( r\right) \right] j_{l}\left( k^{\prime }r\right)
j_{l}\left( kr\right) , \\
V^{-}\left( k,k^{\prime }\right)  &=\frac{2}{\pi }\int r^{2}dr\left[
V\left( r\right) -S\left( r\right) \right] j_{\tilde{l}}\left( k^{\prime
}r\right) j_{\tilde{l}}\left( kr\right) ,
\end{align}%
where $f(k)$ and $g(k)$ are the radial part of Dirac spinor, and $%
j_{l}\left( kr\right) (j_{\tilde{l}}\left( kr)\right) $ are the spherical
Bessel functions of order $l(\tilde{l})$. By turning the integral in Eq.~(%
\ref{Diraceq3}) into a sum over a finite set of points $k_{j}$ and $dk$ with
a set of weights $w_{j}$, it is then transformed into a matrix equation:
\begin{equation}
\sum\limits_{j=1}^{N}\left(
\begin{array}{cc}
A_{ij}^{+} & B_{ij} \\
B_{ij} & A_{ij}^{-}%
\end{array}%
\right) \left(
\begin{array}{c}
f(k_{j}) \\
g(k_{j})%
\end{array}%
\right) =\varepsilon \left(
\begin{array}{c}
f(k_{i}) \\
g(k_{i})%
\end{array}%
\right),   \label{Diraceq4}
\end{equation}%
where $A_{ij}^{\pm }={\pm }M\delta _{ij}+w_{j}k_{j}^{2}V^{\pm }\left(
k_{i},k_{j}\right) $ and $B_{ij}=-k_{i}\delta _{ij}$. In Eq.~(\ref{Diraceq4}), the Hamiltonian matrix is not symmetric. For simplicity in
computation, we symmetrize it by the transformation,
\begin{equation}
\mathfrak{f}(k_{i})=\sqrt{w_{i}}k_{i}f(k_{i}),\quad \mathfrak{g}(k_{i})=\sqrt{w_{i}%
}k_{i}g(k_{i}),
\end{equation}
which gives us a symmetric matrix in the momentum representation as
\begin{equation}
H=\left(
\begin{array}{cc}
\mathcal{A}^{+} & \mathcal{B} \\
\mathcal{B} & \mathcal{A}^{-}%
\end{array}%
\right) ,  \label{DiracMatrix}
\end{equation}%
where $\mathcal{A}_{ij}^{\pm }={\pm }M\delta _{ij}+\sqrt{w_{i}w_{j}}%
k_{i}k_{j}V^{\pm }\left( k_{i},k_{j}\right) $ and $\mathcal{B}_{ij}$ is the
same as $B_{ij}$. So far, to solve the Dirac equation (\ref{Diraceq1}%
) becomes an eigensolution problem of the symmetric matrix (\ref{DiracMatrix}%
). To calculate the symmetric matrix, several key points need to be
clarified. As the integration in Eq.~(\ref{Diraceq3}) is from zero to
infinite, it is necessary to truncate the integration to a large enough
momentum $k_{\text{max}}$. When $k_{\text{max}}$ is fixed, the integration
can be calculated by a sum shown in Eq.~(\ref{Diraceq4}). As a sum with evenly
spaced $dk$ and a constant weight $w_{j}$ converges
slowly, it should not be used. We replace the sum by the Gauss-Legendre
quadrature with a finite grid number $N$, which gives us a $2N\times 2N$
Dirac Hamiltonian matrix (\ref{DiracMatrix}). In the realistic calculations,
we need to choose a proper contour for the momentum integration. From the
scattering theory, we know that the bound states populate on the imaginary
axis in the momentum plane, while the resonances locate at the fourth
quadrant. The contour shown in Fig.~\ref{Fig1}(a) encloses only the bound states. For
the resonant states, the contour needs to be deformed into a complex one
illustrated in Fig.~\ref{Fig1}(b) denoted as L$_{+}$. Using the complex contour L$_{+}$,
one can obtain not only the bound states but also resonant states in the
continuum. As long as the range of contour is large enough, we are able to
get all the concerned resonances. For convenience, we claim this method for
exploring the resonances using the complex momentum representation in the framework
of RMF theory as the RMF-CMR method.

\begin{figure}
\includegraphics[width=8.5cm]{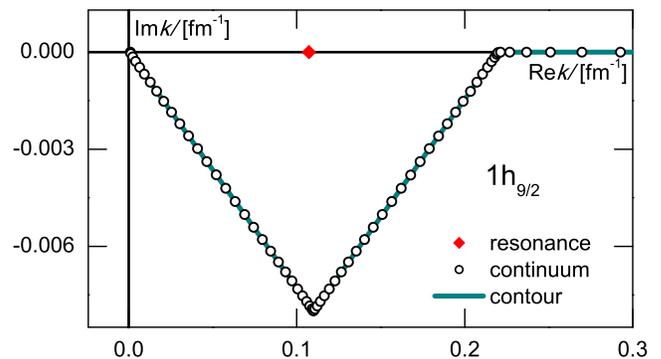}
\caption{(Color online) Single-particle spectra for the state $h_{9/2}$
in the complex $k$-plane in the RMF-CMR calculations with the effective
interaction NL3~\protect\cite{Lal97}. The red filled diamond, black open
circles, and olive solid line represent the $1h_{9/2}$ resonant state, the
continuum, and the contour of integration in the complex momentum plane,
respectively. }
\label{Fig2}
\end{figure}

Using the formalism presented above, we explore the resonances in real
nuclei. By taking the nucleus $^{120}$Sn as an example, we first perform the RMF calculation
with the scalar and vector potentials being obtained. For the resonant
states, the momentum representation is adopted. The Dirac equation is solved
by diagonalizing the matrix (\ref{DiracMatrix}) in the momentum space along
a triangle contour, and the tip of the triangle is placed below the expected
position of the resonance pole. The contour is truncated to
a finite momentum $k_{\text{max}}=3.5$~fm$^{-1}$, which is sufficient for
all the concerned resonances. The grid number of the Gauss-Legendre
quadrature $N=120$ is used for the momentum integration along the contour,
which is enough to ensure the convergence with respect to number of
discretization points. In the practical calculations, the grid number $N$ is
divided into $n$, $n$, and $2n$ used in each segment of the contour,
respectively. For the state $h_{9/2}$, we confine the triangle contour with
the four points $k=0$~fm$^{-1}$, $k=0.11-i0.008$~fm$^{-1}$, $k=0.22$~fm$%
^{-1} $, and $k_{\text{max}}=3.5$~fm$^{-1}$ in the complex $k$-plane. The calculated results are displayed in Fig.~\ref{Fig2},
where we can see that most solutions follow the contour, corresponding to
the non-resonant continuum states. There is one solution that does not lie
on the contour, corresponding to the $1h_{9/2}$ resonance, which is
separated completely from the continuum and exposed clearly in the complex
momentum plane.

\begin{figure}
\includegraphics[width=8.5cm]{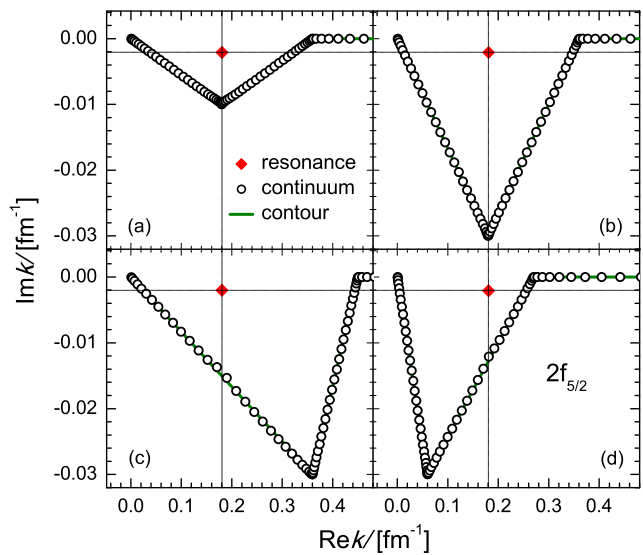}
\caption{(Color online) Same as Fig.~\ref{Fig2}, but for the state $2f_{5/2}$ in
four different contours for the momentum integration. The red filled diamond
in each panel represents the $2f_{5/2}$ resonant state.}
\label{Fig3}
\end{figure}

Although the resonances can be exposed in the complex $k$-plane, we would like to further check whether the present calculations depend on the choice of contour. In
Fig.~\ref{Fig3}, we show the single-particle spectra for the state $f_{5/2}$ in
four different contours. In each panel, one can see a resonant state
exposed clearly in the complex $k$-plane. In comparison with panel (a), the
contour in panel (b) is deeper, and the corresponding continuous spectra drop
down with the contour, while the position of the resonant state $2f_{5/2}$
does not change. Similarly, when the contour moves from left to right or
from right to left, as shown in panels (c) and (d), the continuum follows the
contour, while the resonant state $2f_{5/2}$ keeps at its original position.
These indicate that the physical resonant states obtained by the present
method are indeed independent on the contour.

\begin{figure}
\includegraphics[width=8.5cm]{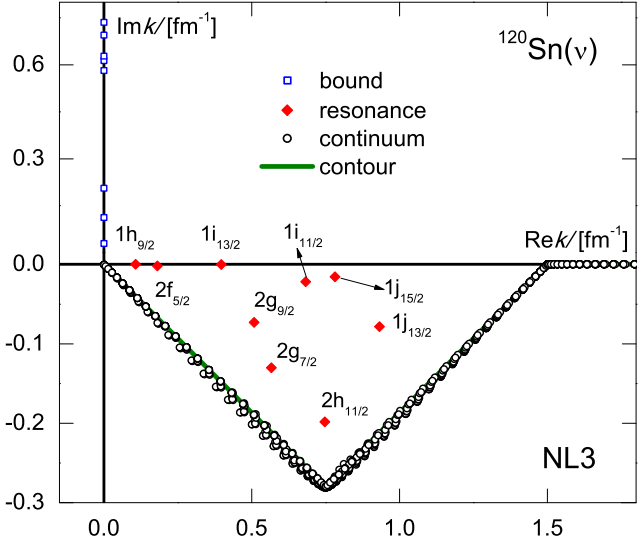}
\caption{(Color online) Single-neutron spectra in $^{120}$Sn in the
RMF-CMR calculations with the interaction NL3. The blue open squares, red
filled diamonds, and black open circles represent the bound states, the
resonant states, and the continuum, respectively. The olive solid line
denotes the contour of integration in the complex momentum plane.}
\label{Fig4}
\end{figure}

As the resonant states are independent on the contour, we can choose a large
enough contour to expose all the concerned resonances. Using the one with
$k=0$~fm$^{-1}$, $k=0.75-i0.28$~fm$^{-1}$, $k=1.5$~fm$^{-1}$, and $k_{\text{%
max}}=3.5$~fm$^{-1}$, the calculated single-neutron spectra in $^{120}$Sn
are shown in Fig.~\ref{Fig4}, where the bound states are exposed on the imaginary
axis, the resonant states are isolated from the continuum in the fourth
quadrant, and the continuum follows the integration contour. Here, we have
observed nine resonant states $1h_{9/2}$, $2f_{5/2}$, $1i_{13/2}$, $%
1i_{11/2} $, $1j_{15/2}$, $1j_{13/2}$, $2g_{9/2}$, $2g_{7/2}$, and $%
2h_{11/2} $. For the resonant states $1h_{9/2}$, $1i_{13/2}$, and $2f_{5/2}$%
, their positions are close to the real $k$-axis, corresponding to the
narrow resonances with smaller widths. For the resonant states $2g_{9/2}$, $%
2g_{7/2}$, and $2h_{11/2}$, they are far away from the real $k$-axis,
corresponding to the broad resonances. Note that these broad resonances have not been
obtained in the RMF-CSM calculations \cite{Guo101} because it requires a
large complex rotation, which leads to the divergence of complex rotation
potential. Similarly, there are also some troubles for exploring these
broad resonance in other methods.
The present RMF-CMR method provides a powerful and efficient pathway to
explore the broad resonances as long as the momentum contour covers the range
of resonance.

\begin{table}
\caption{The calculated energies for the single neutron resonant states in $%
^{120}$Sn varying with the grid number of the Gauss-Legendre quadrature in
the momentum integration. More data are listed in Ref.~[69]. All energies are in units of MeV.}%
\label{Table1}
\begin{tabular}{c|c|c|c|c}
\hline\hline
    & $1h_{9/2}$  & $2f_{5/2}$ 	& $1i_{13/2}$ & $2g_{9/2}$ \\	
$N$	& $E_r$,$E_i$ & $E_r$,$E_i$ & $E_r$,$E_i$ & $E_r$,$E_i$ \\
\hline
60	&	0.239,-2.81E-8	&	0.678,-0.0157	&	3.267,-0.00186	&	5.232,-1.534	\\
80	&	0.239,-2.77E-8	&	0.678,-0.0156	&	3.267,-0.00186	&	5.232,-1.534	\\
100	&	0.239,-2.76E-8	&	0.678,-0.0156	&	3.267,-0.00186	&	5.232,-1.534	\\
120	&	0.239,-2.76E-8	&	0.678,-0.0156	&	3.267,-0.00186	&	5.232,-1.534	\\
140	&	0.239,-2.76E-8	&	0.678,-0.0156	&	3.267,-0.00186	&	5.232,-1.534	\\
160	&	0.239,-2.76E-8	&	0.678,-0.0156	&	3.267,-0.00186	&	5.232,-1.534	\\
\hline\hline
\end{tabular}
\end{table}

When the resonant states are exposed in the complex $k$-plane, we can read
the real and imaginary parts of their wave vectors. We can then extract the resonance parameters like energy and width by the formula
$E_{r}+iE_{i}=E_{r}-i\Gamma /2=\sqrt{k^{2}+M^{2}}-M$. In order to obtain
precise results for the resonance parameters, it is necessary to check the
convergence of the calculated results on the grid number in the
Gauss-Legendre quadrature. The resonance parameters for four resonant
states varying with the grid number are listed in Table~\ref{Table1}, where three
significant digits are reserved in the decimal part. From Table~\ref{Table1}, we can
see that the calculated results are unchange when $N\geqslant 80$ in the
present precision with the $1h_{9/2}$ exception. For the state $1h_{9/2}$,
the tiny difference among the widths should be attributed to the fact that its width is too
small in comparison with the corresponding energy. These imply that we have
obtained the convergent results in the present calculations.

The above discussions indicate that the present method is applicable and efficient for
exploring the resonance. For comparison, the calculated results from
other different bound-state-like methods, RMF-CSM~\cite{Guo101},
RMF-RSM~\cite{LZhang08}, and RMF-ACCC~\cite{Zhang04}, are listed in Table~\ref{Table2} for several narrow single-neutron resonant states in $^{120}$Sn. From
Table~\ref{Table2}, we can see that, in the RMF-CMR calculations with NL3,
the energies and widths for all the available resonant states are comparable to
those obtained by the other methods. The same conclusion can also be
obtained in the RMF-CMR calculations with the effective
interaction PK1~\cite{Long04}, which have not been shown here.

\begin{table}
\caption{Energies and widths of single neutron resonant states for $^{120}$%
Sn in the RMF-CMR calculations in comparison with the RMF-CSM, RMF-RSM, and
RMF-ACCC calculations. Data are in units of MeV.}
\label{Table2}%
\begin{tabular}{c|c|c|c|c}
\hline\hline
& RMF-CMR & RMF-CSM & RMF-RSM & RMF-ACCC \\
$nl_j$ & $E_r, \Gamma$ & $E_r, \Gamma$ & $E_r, \Gamma$ & $E_r, \Gamma$ \\
\hline
$2f_{5/2}$ & 0.678,0.031 & 0.670,0.020 & 0.674,0.030 & 0.685,0.023 \\
$1i_{13/2}$ & 3.267,0.004 & 3.266,0.004 & 3.266,0.004 & 3.262,0.004 \\
$1i_{11/2}$ & 9.607,1.219 & 9.597,1.212 & 9.559,1.205 & 9.60,1.11 \\
$1j_{15/2}$ & 12.584,0.993 & 12.577,0.992 & 12.564,0.973 & 12.60,0.90 \\
\hline\hline
\end{tabular}%
\end{table}

Although the agreeable results are obtained, it is worthwhile to remark the difference in these four different calculations. In the
RMF-ACCC calculations, the resonant states are obtained by extending a bound
state to resonant state, which is effective for the narrow resonances~\cite%
{Zhang04,Guo06,Xu15}, but for the broad resonances the results from the ACCC
calculations are less precise. Compared with the RMF-ACCC, the RMF-RSM is
much simpler. The resonant states can be determined in terms of the
independence of the calculated results on the model parameters. As shown in
Fig.~1 in Ref.~\cite{LZhang08}, there appear the plateau for the resonant
states in the energy surface. For the narrow resonance $1i_{13/2}$, the
plateau is clear, which implies that it is easy to determine the narrow
resonances by the RMF-RSM. Although the CSM is efficient for not only narrow
resonances but also broad resonances, there is the singularity in nuclear
potential with a large complex rotation, which leads that the RMF-CSM is
inapplicable for some broad resonances. Therefore, these four methods are all
effective for the narrow resonances, while only the RMF-CMR method is
applicable and more reliable for the broad resonances.

\begin{figure}
\includegraphics[width=8.5cm]{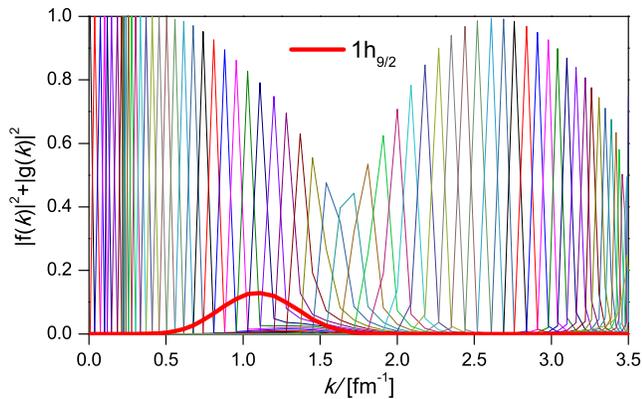}
\caption{(Color online) Radial-momentum probability distributions for
the single-particle states $h_{9/2}$. The thick red line represents the
probability distributions corresponding to the $1h_{9/2}$ resonant state,
while the others are the background of the continuum.}
\label{Fig5}
\end{figure}

Besides the spectra, we have also obtained the wavefunction of Dirac
particle in the momentum space. The radial-momentum probability
distributions (RMPD) for the single-particle states $h_{9/2}$ are drawn in
Fig.~\ref{Fig5}. The RMPD corresponding to the resonance $1h_{9/2}$ is expanded much wider than the surrounding states. The Heisenberg uncertainty principle tells us
that a less well-defined momentum corresponds to a more well-defined
position. Consequently, this state should correspond to a localized
wavefunction, i.e., a wavefunction of resonant state.
Compared with the $1h_{9/2}$, the RMPD for the other states display sharp
peaks at different values of $k$, which correspond to free particles. These indicate that we can also judge the resonance by the wavefunction in the momentum representation.


In summary, we have proposed a new scheme to explore the resonances in the
RMF framework, where the Dirac equation is solved
directly in the complex momentum representation, and the bound and resonant
states are dealt on an equal footing. We have presented the theoretical
formalism and elaborated the numerical details. As an illustrating example,
we have explored the resonances in the nucleus $^{120}$Sn and determined the
corresponding resonance parameters. In comparison with several frequently
used bound-state-like methods, and the agreeable results are obtained. In particular, the present method can expose clearly the resonant states in the
complex momentum plane, and determine precisely the resonance parameters
without imposing unphysical parameters. Also highly remarkable is the present method is
applicable for not only narrow resonances but also broad resonances that are
difficult to be obtained before.

This work was partly supported by the National Natural Science Foundation of
China under Grants No.11575002 and the Key Research Foundation of
Education Ministry of Anhui Province under Grant No. KJ2016A026.

\end{document}